\documentclass[prc,preprint]{revtex4}
\usepackage{graphicx}
\usepackage{amsmath,amssymb}
\usepackage{mathrsfs}
\def\dis{distribution}
\def\pt{p_T}
\def\bq{\begin{eqnarray}}
\def\eq{\end{eqnarray}}
\def\xb{\bar\xi}

\begin{document}

\title{Spectra of Identified Hadrons in Pb-Pb Collisions at LHC}
\author{Rudolph C. Hwa$^1$ and Lilin Zhu$^{1,2}$}
\affiliation
{$^1$Institute of Theoretical Science and Department of
Physics\\ University of Oregon, Eugene, OR 97403-5203, USA\\
$^2$Department of  Physics, Sichuan
University, Chengdu  610064, P.\ R.\ China}
\date{\today}

\begin{abstract}
{The transverse momentum \dis s of identified hadrons produced in Pb-Pb collisions at the Large Hadron Collider (LHC) are studied in the low and intermediate range for $\pt<5$ GeV/c. All four spectra ($\pi, K, p, \Lambda$) can be well reproduced in the recombination model based on a common thermal parton \dis\ of light and strange quarks and on shower partons emitted in hard and semihard jets. Two essential parameters are adjusted to fit the data, one being the inverse slope of the thermal \dis\ and the other revealing the degree of momentum degradation in the medium. Various combinations of thermal and shower parton components are calculated, showing the dominance of minijets. The effect of minijets is to produce harder baryons than mesons resulting in their ratio to peak at around $\pt\sim 3$ GeV/c. Substantial portion of the jet energy is found to be lost to the dense medium before the partons emerge at the surface to undergo hadronization by recombination.}
\end{abstract}
\maketitle

\section{Introduction}

Recent data from ALICE on Pb-Pb collisions at $\sqrt{s_{NN}}=2.76$ TeV provide the first glimpse of the transverse-momentum ($p_T$) distributions of
identified hadrons produced at the Large Hadron Collider (LHC) \cite{mf}.  Although the $p_T$ ranges are from 0 to 3 GeV/c only for $\pi$ and $p$, and up to 5 GeV/c for $K^0$ and $\Lambda^0$, which are very low by LHC standards, there are already features that are unexpected by extrapolation from RHIC \cite{fa}.  The proton distribution seems to be pushed harder to higher $p_T$ so that the $p/\pi^+$ ratio, as well as the $\bar p/\pi^-$ ratio, continue to rise up to the highest detected momentum at $p_T \approx 3$ GeV/c.  The $\Lambda/K$ ratio does show a well-developed peak at $p_T = 3$ GeV/c, and the magnitude of the ratio increases with centrality, exceeding 1.5 at 0-5\% centrality.  Since no model-calculation has been cited in Refs. \cite{mf,fa} to indicate the existence of any satisfactory explanation of the data, it is important to investigate this problem and develop a model that can accommodate all aspects of the $p_T$ distributions for $\pi, K, p$ and $\Lambda$ production up to 5 GeV/c.

The region of low and intermediate transverse momenta, $0<\pt<5$ GeV/c, is difficult for theoretical treatment that can claim validity throughout the whole range. It is too low for perturbative QCD and includes $\pt$ too high for conventional hydrodynamics. In our study here we focus only on the hadronization process for which we use the recombination model that has been shown to be effective in interpreting the data at intermediate $\pt$ \cite{hwa}. The input involves thermal and shower parton \dis s at late time that are phenomenological in the sense of having some parameters not determined from first principles. However, our approach has the advantage of making transparent the common origin of the various components that contribute to the spectra of different hadrons produced.
Since the recombination model has been successful in explaining the large baryon/meson ratio found at RHIC \cite{hy, gkl, fmnb},  it is sensible to extend that approach here in examining its applicability at LHC.  It is generally recognized that at high energy in the TeV regime minijets are copiously produced from semihard scatterings.  The density of shower partons from such minijets \cite{hy1, hy2} is therefore expected to be much higher than at RHIC.  On the other hand, thermal partons at hadronization should not depend sensitively on the collision energy, since the hadronization of a system of higher initial density occurs later when the thermal medium can expand to a lower local density in order for confinement to take place.  Thus the mixture of shower and thermal partons at late time at LHC should be very different from that at RHIC.  Since shower partons are harder than thermal partons, we therefore expect the produced hadrons to be also harder --- more so for the baryons than mesons due to their larger number of constituents in the recombination model.  This is qualitatively the short explanation for why the ALICE data exhibit harder baryon spectra compared to the RHIC data.  The essence of this work is to carry out the investigation quantitatively.

The dense partonic medium created at LHC can lead to severe attenuating effects on semihard partons that traverse the medium.  One can focus on the various mechanisms of energy loss in QCD, as in \cite{dks}, but such studies of medium effects do not lend themselves readily to the analysis of the hadronization process outside the medium, which is our goal.  A template for a phenomenological description of the medium effects has been found in Ref. \cite{hy3} that takes geometrical properties of Au-Au collisions at RHIC into consideration.  We shall use the general form of that study for LHC, but not  its details.  With the assumption of the dominance of gluon jets for the determination of all shower partons, the effect of jet quenching is common to all flavors of shower partons that undergo hadron formation outside the medium.  Thus the central issues in this work are on the various ways in which thermal and shower partons can recombine at $p_T < 5$ GeV/c in Pb-Pb collisions at 2.76 TeV.  To achieve a satisfactory description of the spectra of all identified species is our immediate goal so that the study of hadron production at higher $p_T$ can follow at a later stage on a firm footing.

\section{Meson and Baryon Inclusive $p_T$ Distributions}
Let us first collect here the basic equations in the recombination model developed previously \cite {hy2, hy3}.  For the invariant $p_T$ distributions of meson and baryons averaged over $\eta$ and $\phi$ at midrapidity, we may use the one-dimensional form
\begin{eqnarray}
p^0{dN^M\over dp_T}&=&\int {dp_1\over p_1}{dp_2\over p_2} F_{q_1\bar q_2}(p_1,p_2) R_{q_1,\bar q_2}^M(p_1,p_2,p_T) ,  \label{1} \\
p^0{dN^B\over dp_T}&=&\int \left[\prod_{i=1}^3 {dp_i\over p_i} \right] F_{q_1q_2q_3}(p_1,p_2,p_3) {R}_{q_1,q_2,q_3}^B(p_1,p_2,p_3,p_T) ,    \label{2}
\end{eqnarray}
where $p_i$ is the transverse momentum (with the subscript $T$ omitted) of one of the recombining quarks $q_i$, whose evolution from a parton emitted at the medium surface to a valon (the constituent quark of the hadron to be formed) is a dressing process that preserves its momentum \cite{hwa1}.  The recombination functions (RFs), $R^{M,B}$, that include the effects of dressing and hadronic structure, have been determined previously and will be given below.  We note here that our $dN^h/dp_T$ should be identified with the experimental $dN/2\pi d\eta dp_T$, which integrates over all $\phi$, while our quantity is defined as averages over $\phi$ and $\eta$.  The parton distributions can be partitioned into various components represented symbolically by
\begin{eqnarray}
F_{q_1\bar q_2}&=&{\cal TT+TS+SS} ,  \label{3} \\
F_{q_1q_2q_3}&=&{\cal TTT+TTS+TSS+SSS},     \label{4}
\end{eqnarray}
where $\cal T$ and $\cal S$ are the invariant distributions of thermal and shower partons, respectively.  There can be many terms within each of the components shown above.

It should be noted that the nature of time evolution of the system is not specified in Eqs.\ (\ref{1}) - (\ref{4}).  The recombination model describes hadronization at late time.  Thus $\cal T$ and $\cal S$ distributions are model inputs at that time that are appropriate for particular collision systems.  The thermal distribution should not depend sensitively on the collision energy, 
since the density of the thermal system must be reduced by expansion to around the same level at any collision energy in order for hadronization to occur. Thus the thermal parton distribution that we adopt has the same form as before \cite{hy2}
\begin{eqnarray}
{\cal T}(p_1) = p_1{dN_q^T\over dp_1}=C p_1e^{-p_1/T},     \label{5}
\end{eqnarray}
where $C$ has the dimension of inverse momentum.
The prefactor $p_1$ is necessary to yield the exponential behavior for the thermal component of the hadronic \dis\ $dN_h/\pt d\pt$, as we shall see after specifying the RF. [See Eq.\ (\ref{31}) below.] 
On the other hand, the properties of shower \dis\ do depend strongly on the collision energy, not only because of the increased rate of creation of hard partons, but also due to the quenching effect of the denser medium. With $\xi$ used as a parameter that describes an aspect of the momentum degradation in the medium, $F_i(q,\xi)$ is the hard or semi-hard parton distribution at the surface of the medium to be discussed in detail below.  For now we just show
the shower  distribution as 
\begin{eqnarray}
{\cal S}(p_2,\xi)=\int {dq\over q}\sum_i F_i(q,\xi) S_i(p_2/q),  \label{6}
\end{eqnarray}
where  $S_i(z)$ is the shower parton distribution (SPD) in a jet of type $i$ with momentum fraction $z$.    In Refs.\ \cite{hy1,hy4} SPD is determined by regarding the fragmentation function (FF), $D_i(x)$, as being the recombination product of two shower partons
\begin{eqnarray}
xD_i^{\pi}(x) = \int {dx_1\over x_1}{dx_2\over x_2} \left \{S_i^q(x_1), S_i^{\bar q}\left({x_2\over 1-x_1}\right)\right \}R_{q{\bar q}}^{\pi}(x_1,x_2,x),
\label{7}
\end{eqnarray}
where the curly brackets denote symmetrization of the leading parton momentum fractions $x_1$ and $x_2$.

Using Eqs.\ (\ref{3})-(\ref{6}) in (\ref{1}), we obtain for pion production
\begin{eqnarray}
p^0{dN_{\pi}^{TT}\over dp_T} &=& C^2 \int dp_1dp_2e^{-(p_1+p_2)/T} R^{\pi}(p_1,p_2,p_T),   \label{8} \\
p^0{dN_{\pi}^{TS}\over dp_T} &=& \int d\xi P(\xi, \phi, b) \int {dq\over q} \sum_i F_i(q,\xi) \widehat{TS}(q,p_T),     \label{9} \\
p^0{dN_{\pi}^{SS}\over dp_T} &=& \int d\xi P(\xi, \phi, b) \int {dq\over q} \sum_i F_i(q,\xi) \widehat{SS}(q,p_T),     \label{10}
\end{eqnarray}
where
\begin{eqnarray}
\widehat{TS}(q,p_T) &=& \int {dp_1\over p_1}{dp_2\over p_2} {\cal T}^{\bar q}(p_1)S_i^q\left({p_2\over q}\right){R}^{\pi}_{q\bar q}(p_1,p_2,p_T),     \label{11} \\
\widehat{SS}(q,p_T) &=& \int {dp_1\over p_1}{dp_2\over p_2} \left\{ S_i^{\bar q}\left({p_1\over q}\right), S_i^q\left({p_2\over q-p_1}\right)\right\}
{R}^{\pi}_{q\bar q}(p_1,p_2,p_T) \nonumber \\
&=& {p_T\over q} D_i^{\pi}(p_T/q) .    \label{12}
\end{eqnarray}
 $P(\xi,\phi,b)$ is the probability for the dynamical path length to be $\xi$ for a path at angle $\phi$ initiated at $(x_0,y_0)$, weighted by the nuclear overlap function, and integrated over all $(x_0,y_0)$.  Geometrically, $\xi$ depends on where the trajectory is between the center and the periphery of the overlap, and dynamically it depends on the energy loss along the path, as will be discussed in more detail in the next section.

For notational brevity we define
\begin{eqnarray}
\bar F_i(q,\bar \xi) = \int d\xi P(\xi,\phi,b)F_i(q,\xi)    \label{13}
\end{eqnarray}
where $\bar\xi$ depends on $\phi$ and $b$, and get from Eqs.\ (\ref{9}) and (\ref{11})
\begin{eqnarray}
{dN_{\pi}^{TS}\over p_Tdp_T} = {2\over p_T^2} \int {dp_1\over p_1}{dp_2\over p_2} {\cal T}^{\bar q}(p_1){\cal S}^q(p_2,\bar \xi){R}^{\pi}_{q\bar q}(p_1,p_2,p_T),     \label{14}
\end{eqnarray}
where
\begin{eqnarray}
{\cal S}^q(p_2,\bar \xi)=\int {dq\over q}\sum_i \bar F_i(q,\bar \xi) S_i^q(p_2/q)  \label{15}
\end{eqnarray}
that follows obviously from Eqs.\ (\ref{6}) and (\ref{13}), except to emphasize that the medium effect is in $\bar F_i(q,\bar\xi)$, not in $S_i^q(p_2/q)$. That is, the FF in Eq.\ (\ref{7}) is for partons outside the medium and is therefore not modified by it. The factor of 2 in Eq.\ (\ref{14}) arises from the two distinguishable components of ${\cal T}^u(1){\cal S}^{\bar d}(2) + {\cal T}^{\bar d}(1){\cal S}^u(2)$.  Such a factor is absent in the $TT$ term because ${\cal T}^u(1){\cal T}^{\bar d}(2)$ and ${\cal T}^{\bar d}(1){\cal T}^u(2)$ are indistinguishable in a thermalized medium of $q$ and $\bar q$.  For $K^+$ production we have for $TS$ two terms of $u\bar s$, so we obtain
\begin{eqnarray}
{dN_K^{TS}\over p_Tdp_T} = {1\over p^0p_T} \int {dp_1\over p_1}{dp_2\over p_2} \left[{\cal T}^q(p_1){\cal S}^s(p_2,\bar \xi) + {\cal T}^s(p_2){\cal S}^q(p_1,\bar \xi)\right]{R}^K(p_1,p_2,p_T),     \label{16}
\end{eqnarray}
where ${\cal S}^s$ is as defined in Eq.\ (\ref{15}), but with $S_i^q$ replaced by $S_i^s$.  Because of the mass difference between the two constituents of $K$, $R^K(p_1,p_2,p_T)$ is not symmetric under the interchange of $p_1$ and $p_2$; moreover, the two terms in the square brackets are different.  Since we consider only hadron production at mid-rapidity here, we may approximate $p^0$ by the transverse mass $m_T$, where
\begin{eqnarray}
m_T = (m_h^2 + p_T^2)^{1/2}.     \label{17}
\end{eqnarray}
For $h=K, p$ and $\Lambda$, it can be significantly different from $p_T$ when $p_T$ is small, and can cause the low-$p_T$ spectra to deviate from the exponential behavior of the pion spectrum.

The recombination of shower partons from the same jet is equivalent to fragmentation, so we have
\begin{eqnarray}
{dN_M^{SS}\over p_Tdp_T} = {1\over m_Tp_T} \int {dq\over q}\sum_i \bar F_i(q,\bar \xi){p_T\over q}D_i^M \left({p_T\over q} \right).     \label{18}
\end{eqnarray}
We shall denote it as $(SS)^{1j}$.  It is also possible for shower partons from adjacent minijets to recombine; we shall refer to such processes as $(SS)^{2j}$.

For baryon production the parton distribution shown in Eq.\ (\ref{4}) has many components that can be expressed more explicitly, though still symbolically condensed, as
\begin{eqnarray}
F_{q_1q_2q_3} = {\cal TTT + TTS + T(SS)}^{1j} + {\cal (SSS)}^{1j} + {\cal T(SS)}^{2j} +{\cal  ((SS)}^{1j}{\cal S)}^{2j} + {\cal (SSS)}^{3j}   \label{19}
\end{eqnarray}
in obvious notation except for $(({\cal SS})^{1j}{\cal S})^{2j}$ which means that two shower-partons are from 1-jet and the third one from a second jet.  We have found that all terms from multi-jets are small for $p_T < 5$ GeV/c, so in this paper we ignore 2j and 3j and omit the specification 1j, since only 1-jet is considered.  If we use the simplified notation to abbreviate Eqs.\ (\ref{14}) and (\ref{16}) as
\begin{eqnarray}
(TS)_{\pi} = 2T_qS_q, \quad (TS)_K = T_qS_s + T_sS_q,     \label{20}
\end{eqnarray}
then in a similar way we shorten the expression for $p$ and $\Lambda$ as
\begin{eqnarray}
(TTS)_p &=& T_qT_q(S_u+S_d),      \label{21} \\
(T(SS))_p &=& T_q(S_uS_d + S_uS_u),     \label{22} \\
(SSS)_p &=& S_uS_uS_d, \\     \label{23}
(TTS)_{\Lambda} &=& T_q(T_qS_s + 2T_sS_q),   \label{24} \\  
(T(SS))_{\Lambda} &=& 2T_qS_qS_s + T_sS_uS_d,    \label{25} \\ 
(SSS)_{\Lambda} &=& S_uS_dS_s,     \label{26}
\end{eqnarray}

The recombination functions (RFs)  have been determined from the study of hadronic structure \cite{hy2,hwa1,hy5}.  They are given below in terms of $y_i$, which is the momentum fraction of a valon (which plays the role in the scattering problem as a constituent quark does in the bound-state problem) relative to the $p_T$ of the produced hadron; i.e., $y_i=p_i/p_T$,
\begin{eqnarray}
{R}^{\pi}(y_1,y_2) &=& y_1y_2\delta(y_1+y_2-1),   \label{27}  \\  
{R}^K(y_1,y_2) &=& B^{-1}(a+1, b+1)y_1^{a+1}y_2^{b+1}\delta(y_1+y_2-1),    \label{28} \\ 
{R}^B(y_1,y_2,y_3) &=& g^{B}_{st}N_B(y_1y_2)^{\alpha +1}y_3^{\beta + 1}\delta(y_1+y_2+y_3 - 1),     \label{29}
\end{eqnarray}
where
\begin{eqnarray}
N_B = [B(\alpha+1, \alpha+\beta+2) B(\alpha+1,\beta+1)]^{-1},     \label{30}
\end{eqnarray}
and $g^{B}_{st}$ is a statistical factor.  From previous studies it has been determined that $a=1, b=2$ \cite{hy6}, $\alpha=1.75, \beta=1.05$ for proton \cite{hy5} and $\alpha=1, \beta=2$ for $\Lambda$ \cite{hy6}.  
For pion $a=b=0$ because its mass is especially low due to its being a Goldstone boson, so tight binding of the constituent quarks leads to broad \dis\ in $y_i$.

An immediate consequence of the momentum-conserving $\delta$-functions in the RFs is the simplification of the hadronic \dis s. We illustrate that by writing out explicitly the pion 
\dis s. Substituting Eq.\ (\ref{27}) in (\ref{8}) and (\ref{14}), we get 
\begin{eqnarray}
{dN_\pi^{TT}\over \pt d\pt}&=&{C^2\over 6} e^{-\pt/T} ,   \label{31} \\
{dN_\pi^{TS}\over \pt d\pt}&=&{2C\over \pt^3} \int_0^{\pt} dp_1 p_1 e^{-p_1/T} {\cal S}^q(\pt-p_1,\bar\xi) ,   \label{32} 
\end{eqnarray}
where the pion mass in $p^0=m_T$ is neglected. The $SS$ component can be written in terms of FF as in Eq.\ (\ref{18}), for which the properties of RF are already used in relating $S$ to $D$ in Eq.\ (\ref{7}). Similar equations as the two above can be exhibited for other hadrons, as done in Appendix A. Here we show the thermal component of proton
\begin{eqnarray}
{dN_p^{TTT}\over \pt d\pt}=N'_p C^3 {\pt^2\over m_T} e^{-\pt/T} ,   \label{33} 
\end{eqnarray} 
where $N'_p=g^{p}_{st}N_B B(\alpha+2,\beta+2)B(\alpha+2,\alpha+\beta+4)$, so that we can emphasize, by comparing Eqs.\ (\ref{31}) and (\ref{33}), that pion and proton have the  same exponential factors; however, the latter has a prefactor $\pt^2/m_T$ that arises from the kinematics of recombination, causing the $\pi$ and $p$ spectra to differ at low $\pt$. Note that the inverse slopes $T$ for both spectra are the same.

The objective in this section  to express the hadronic spectra in terms of the parton \dis s is now accomplished. In the following it is then only necessary for us to focus  on the latter in order to calculate the former.

\section{THERMAL AND SHOWER PARTONS}

As noted at the end of the preceding section the thermal \dis s of $p$ and $\pi$ have the same inverse slope $T$, inherited from the thermal partons, whose invariant \dis\ is given in Eq.\ (\ref{5}). This is the characteristic property of recombination that deviates from the hydrodynamical description where $T_{\rm slope}$ depends on the hadron mass due to flow effect \cite{ph}.  There is apparent disparity between the $p$ and $\pi$ spectra in the data;  nevertheless, because of the prefactor $p_T^2/m_T$ for proton and of the dominant resonance contribution to pion for $p_T<1$ GeV/c, it does not mean that it is impossible to have a good description of both spectra using one value of $T$, as will be shown below.

We now extend the universality of the thermal parton \dis\ to include the strange sector in hadron production at LHC. In the TeV realm of collision energies it is expected that the light and strange quarks are fully equilibrated before hadronization. If that is so, then the same thermal \dis\ given in Eq.\ (\ref{5}) should be valid for $u, d$ and $s$ quarks, the consequence of which can readily be checked in our model calculation. We note that the issue is not one that concerns hadronic masses, since the system under discussion consists only of thermal partons in local equilibrium. More specifically, before the recombination mechanism is applied, we assume that all gluons are converted to $u, d$ and $s$ quarks and their antiquarks, which form the initial states of specific channels for hadronization \cite{hwa1}. Thermal gluons do not fragment in the usual sense. Even fragmentation of gluon jets is treated by conversion to $q$ and $\bar q$ before recombination.

For shower partons let us begin by first summarizing the subject of momentum degradation treated in Ref.\ \cite{hy3}. At RHIC the $p_T, \phi$ and centrality ($c$) dependencies of the nuclear modification factor $R_{AA}(p_T,\phi,c)$ impose stringent constraints on the dynamical process of energy loss. The problem is complicated because of both the geometrical and dynamical aspects of the description of the parton's traversal through a non-uniform medium. For a given point $(x_0,y_0)$ in the transverse plane where a hard parton is created in the initial system in a collision at impact parameter $b$, one has to calculate the geometrical path length $\ell(x_0,y_0,\phi,b)$ of a path at angle $\phi$ and then the medium effect along that path. As the system expands, $\ell$ becomes longer, but the local density becomes lower, so those compensating effects on the net energy loss result in a dynamical path length $\xi$ that can be related to $\ell$ through an undetermined parameter but without a time-dependent transport description. Upon averaging over all creation points $(x_0,y_0)$, one arrives at a probability \dis\ on $\xi$, denoted by $P(\xi,\phi,b)$, that relates the average parton \dis\ $\bar F_i(q,\phi,b)$ with momentum $q$ at the medium surface to the \dis, $F_i(q,\xi)$, with a definite $\xi$ by a weighted average
\begin{eqnarray}
\bar F_i(q,\phi,b)&=&\int d\xi P(\xi,\phi,b) F_i(q,\xi) .     \label{34}
\end{eqnarray}
This is a general relationship with details all contained in $P(\xi,\phi,b)$. The parton \dis\ $F_i(q,\xi)$
 is related in turn to the distribution $f_i(k)$ at the point of creation by
\begin{eqnarray}
F_i(q,\xi)&=&\int dk k f_i(k) G(k,q,\xi) ,  \label{35}
\end{eqnarray}
where $f_i(k)$ is the parton density in the phase space $kdk$. 
  $G(k,q,\xi)$ is the momentum degradation function from $k$ to $q$:
\begin{eqnarray}
G(k,q,\xi)=q\delta(q-ke^{-\xi}).  \label{36}
\end{eqnarray}

In \cite{hy3} the  RHIC data \cite{sa} on $R_{AA}(\pt,\phi,c)$ are used to constrain the properties of $P(\xi,\phi,b)$. It is found that $P(\xi,\phi,b)$ can be expressed in some scaling form involving the scaling variable $z=\xi/\bar\xi$, where $\bar\xi(\phi,b)=\int d\xi\, \xi P(\xi,\phi,b)$. For our purpose here, let us not repeat the details of that study for Pb-Pb collisions at LHC, especially when we restrict our attention in this work to only the data at most central collisions. We circumvent the complications by deriving a simple parametric form as follows. From Eqs.\ (\ref{13}), (\ref{34})-(\ref{36}) we have for the average parton \dis\ at the surface 
\begin{eqnarray}
\bar F_i(q,\bar \xi) = \int d\xi P(\xi,\phi,b) q^2e^{2\xi}f_i(qe^{\xi}) ,     \label{37}
\end{eqnarray}
which we parametrize for centrality $0<c<0.05$ in the form
\begin{eqnarray}
\bar F_i(q,\kappa) = k(q)^2 f_i(k(q)) ,   \qquad k(q)=\kappa q .    \label{38}
\end{eqnarray}
If the details of $P(\xi,\phi,b)$ discussed in \cite{hy3} are used, it can be shown that $\bar F_i(q,\bar\xi)$ in Eq.\ (\ref{37}) for a relevant value of $\bar\xi$ can be reproduced very closely by Eq.\ (\ref{38}) for a corresponding value of $\kappa$. Since in either case an unknown parameter is needed to represent the effect of momentum degradation, we choose the latter expression that does not rely on the details of $P(\xi,\phi,b)$, and treat  $\kappa$ as the key parameter to fit the data of all hadronic spectra. The significance of $\kappa$ is clearly the average momentum fraction, $\kappa^{-1}$, that a parton retains upon traversing the medium.

\section{PARAMETRIZATION}

For the thermal parton \dis\ we use the form in Eq.\ (\ref{5}) and put 
\bq
C=23.2\ ({\rm GeV/c})^{-1} \label{39}
\eq
that is the value determined at RHIC  \cite{hy2}. We shall let $T$ be adjustable to fit the LHC data at low $\pt$. Since our aim is to reproduce the $\pt$ \dis s of all identified particles ($\pi, K, p, \Lambda$), the use of one parameter $T$ for all thermal partons is not only economical, but would indeed be remarkable, if achieved. Unlike in hydrodynamical studies that consider mass-dependent flow effect, our approach incorporates the effects of minijets on the hadronic spectra at low $\pt$, allowing the recombination processes to determine the similar or dissimilar behaviors of different hadronic species.

The parton distributions,  $f_i(k)$, at creation  have been determined in Ref.\ \cite {sgf} for Au-Au collision
at 0.2 TeV and for Pb-Pb collision at 5.5 TeV.  The form used is
\begin{eqnarray}
f_i(k) = K{A\over (1 + k/B)^n},     \label{40}
\end{eqnarray}
where $K=2.5$, $A, B$ and $n$ are tabulated for various quarks and gluon.  For $\sqrt{s_{NN}}=2.76$ TeV we use logarithmic interpolation between the
two energies for $\ell nA, B$ and $n$.  More specifically, for gluons we use
\begin{eqnarray}
A = 6.2 \times 10^4 \ {\rm GeV}^{-2}, \quad  B = 0.98 \ {\rm GeV}, \quad  n=6.22.     \label{41}
\end{eqnarray}
Other parameters can similarly be obtained.  The quark contributions are found to be significantly less than the gluon contribution. To make our calculation transparent, we shall assume the dominance of gluon jets and let the summation over parton species in all hard parton terms be represented by a factor $\sigma$ times the contributions from the gluon jets only, which we shall calculate explicitly using Eqs.\ (\ref{40}) and (\ref{41}) for gluon creation at $k>3$ GeV/c. That is, we approximate $\sum_i F_i$ by $\sigma F_g$ and shall use $\sigma=1.2$ in the calculation, since the quark jets amount to roughly 20\% of the gluon jet.

Since the parton distribution $\bar F_i(q,\kappa)$ is not reliable for $q < 3$ GeV/c, its contribution at low $q$ is cut off by a smooth function
which we take to be
\begin{eqnarray}
g(q)=\left[1+e^{(3.5-q)/0.5}\right]^{-1}.     \label{42}
\end{eqnarray}
With this cut-off factor we may then let the $q$ integrals in all equations, such as in Eqs.\ (\ref{6}) and (\ref{15}), to start at $q=0$. 
Shower parton with momentum $p_j$ has a momentum fraction $x=p_j/q$ that peaks at small $x$; consequently, even for $q > 3$ GeV/c, the density of
shower partons at $p_j < 1$ GeV/c can become very high at LHC.  At low $p_T$ we know that thermal partons are important.  How to separate the shower partons from the thermal partons at low $p_j$ is not very well defined.  The energy lost by semihard partons as they traverse the medium can enhance the thermal motion of the soft partons in the vicinities of their trajectories, thereby contributing to a component of the thermal partons that are
intimately related to the soft component of the shower partons.  
The effective $T$ that we shall determine includes enhanced thermal partons due to the energy loss of the semihard partons. Because of that effect the distributions of shower partons that are to be determined from the FF according to Eq.\ (\ref{7}) must be modified at low $x$  so that ${\cal S}(p_1)$ will not exceed ${\cal T}(p_1)$.  That modification is discussed in detail in Appendix B.
The phenomenological basis that supports our approach to the problem is the success in treating the pion and proton spectra at  RHIC from a common partonic distribution characterized by a universal $T$ \cite{hy2, hz, ssa}. Here we apply the same idea to the problem at LHC. 

To summarize, we have two basic parameters to adjust to fit a large collection of data from LHC. Those parameters are $T$ and $\kappa$ that characterize the thermal medium and its quenching effects on minijets. Due to the copious production of minijets at LHC, we expect that  the relative magnitudes of the parton \dis s ${\cal T}$ and ${\cal S}$ will be different from those at RHIC. They contribute to hadronic \dis s in the intermediate range $1<\pt<5$ GeV/c that is hard to quantify in other approaches. Thus our phenomenological treatment that employs two free parameters is a worthwhile endeavor, provided that all four hadronic spectra can be reproduced.

 \begin{figure}[tbph]
\centering
\includegraphics[width=0.6\textwidth]{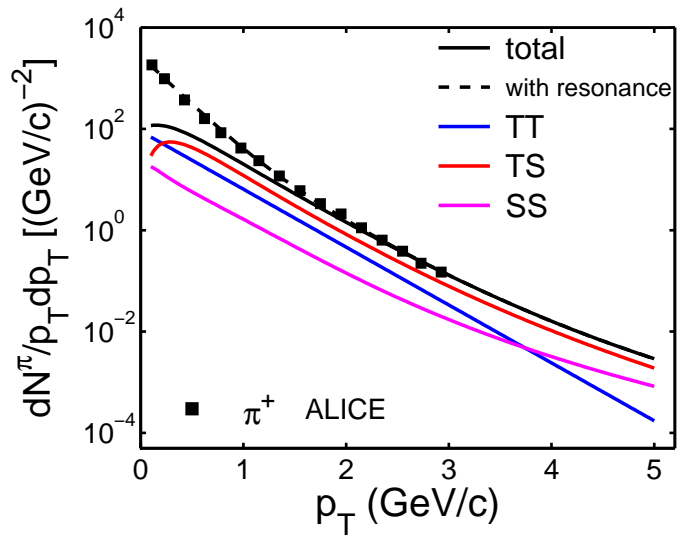}
\caption{(Color online) Transverse momentum \dis\ of pion produced in Pb-Pb collision at $\sqrt {s_{NN}}=2.76$ TeV. Data are from Ref.\ \cite{mf} for centrality 0-5\%. The black curve is the sum of the components calculated for TT (blue), TS (red), SS (purple) recombination. The dashed line is a fit to the data, whose difference from the solid is attributed to resonance decay.}
\end{figure}
 \begin{figure}[tbph]
\centering
\includegraphics[width=0.6\textwidth]{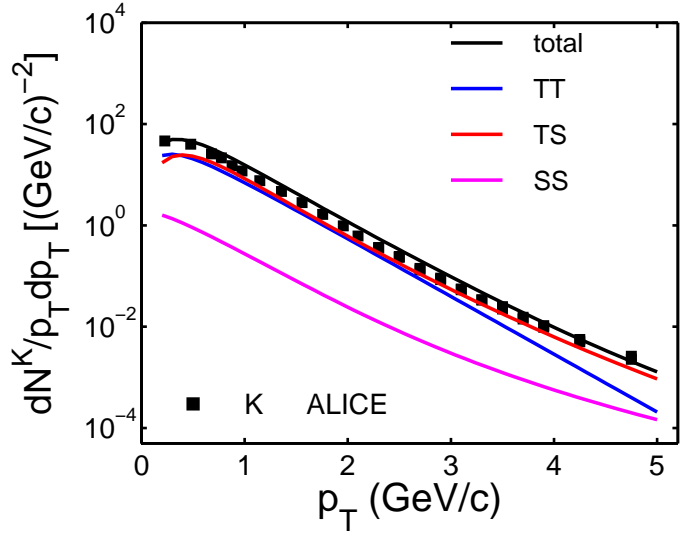}
\caption{(Color online) 
Transverse momentum \dis\ of $K$ produced in Pb-Pb collision at $\sqrt {s_{NN}}=2.76$ TeV. 
Data are from Ref.\ \cite{mf} for centrality 0-5\%. The data points in the region $\pt<2$ GeV/c are for $K^+$, while those for $\pt>2$ are obtained from the ratio $\Lambda^0/K^0_s$ and the $\Lambda^0$ spectrum in \cite{mf,ib}. The black curve is the sum of the components calculated for TT (blue), TS (red), SS (purple) recombination.}
\end{figure}
\begin{figure}[tbph]
\centering
\includegraphics[width=0.6\textwidth]{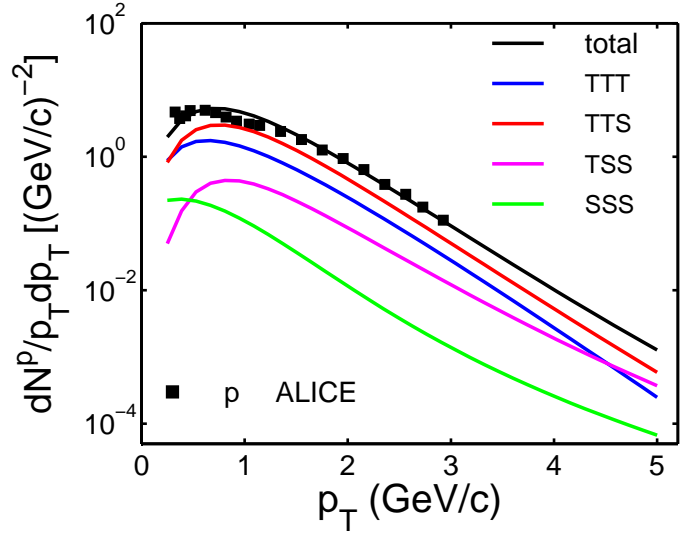}
\caption{(Color online) Transverse momentum \dis\ of proton produced in Pb-Pb collision at $\sqrt {s_{NN}}=2.76$ TeV. Data are from Ref.\ \cite{mf} for centrality 0-5\%. The black curve is the sum of the components calculated for TTT (blue), TTS (red), TSS (purple), SSS (green) recombination.}
\end{figure}
\begin{figure}[tbph]
\centering
\includegraphics[width=0.6\textwidth]{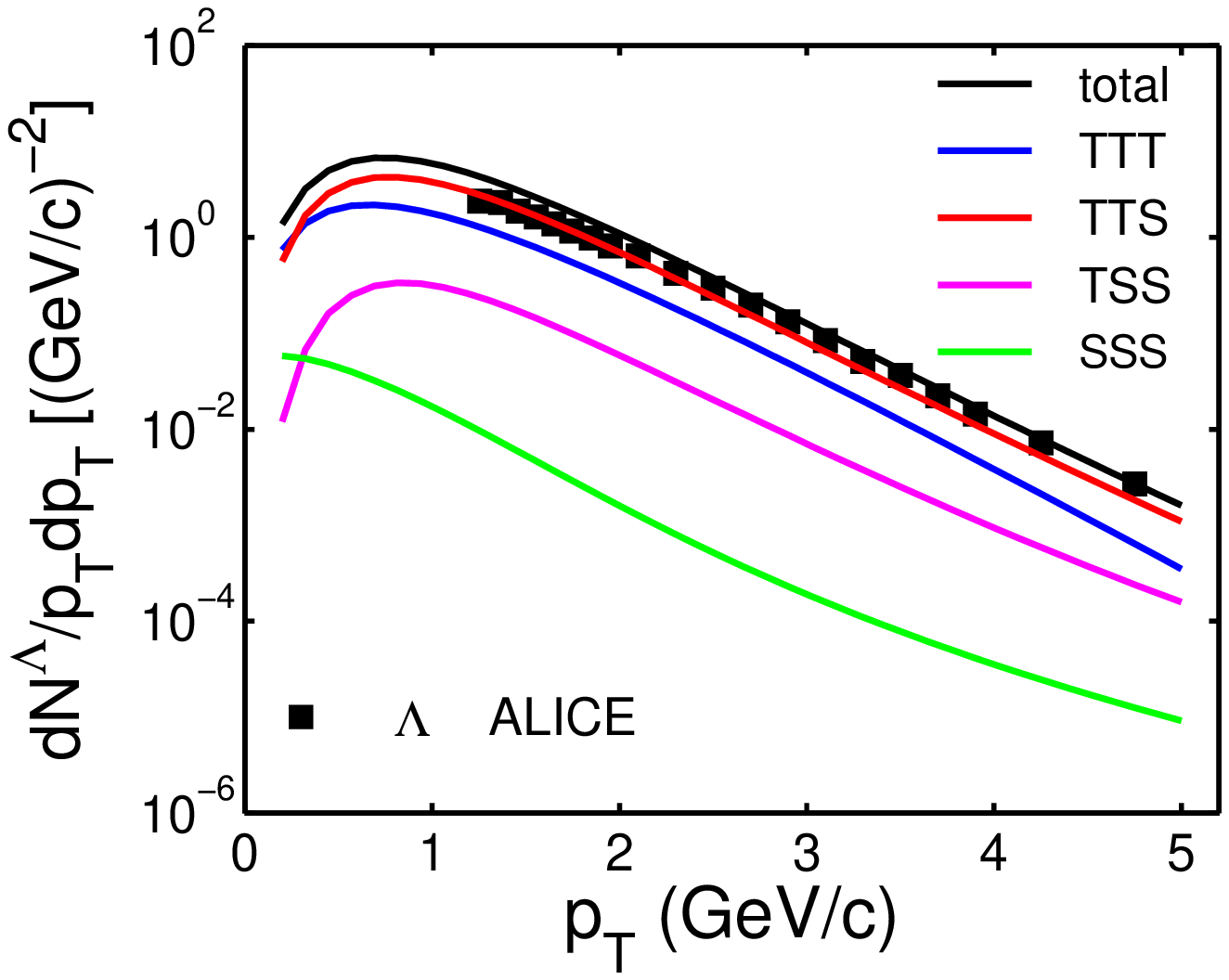}
\caption{(Color online) Transverse momentum \dis\ of $\Lambda$ produced in Pb-Pb collision at $\sqrt {s_{NN}}=2.76$ TeV. Data are from Ref.\ \cite{mf} for centrality 0-5\%. The lines are color-coded as in Fig.\ 3.}
\end{figure}

\section{RESULTS}

We calculate  the four hadronic spectra using equations given in Appendix A. 
In all those equations parton \dis s involving the parameter $\bar\xi$ are to be replaced by the \dis s in terms of $\kappa$, as we have discussed in going from Eq.\ (\ref{37}) to (\ref{38}).
We have varied $T$ and $\kappa$ to achieve the best fits of the data reported by ALICE for central (0-5\%) Pb-Pb collisions at 2.76 TeV \cite{mf}. 
The results are shown in Figs.\ 1-4.
The values of those parameters are 
\bq
T=0.38\ {\rm GeV},  \qquad\quad \kappa=2.6.   \label{43} 
\eq
The color code for the various lines representing the different components are as follows: TT and TTT in blue, TS and TTS in red, SS and TSS in purple, SSS in green, and the total in black. The agreement between the black lines and the data points for $K, p$ and $\Lambda$
 in Figs.\ 2-4 are evidently very good throughout the whole region where data exist.
 
In the case of pion the calculated total falls below the data for $\pt<1.5$ GeV/c. 
The dashed (black) line is drawn to fit that region and represents
 the extra contribution from the decay of resonances. Such a component cannot at present be calculated in this formalism, because the RFs have been derived from the hadronic structures of the lowest bound states. Resonances involve orbital excitations for which the RFs have not been investigated. 
 The dominance of resonance production at low $\pt$ is a known fact even for meson-proton collisions at ${\sqrt s}=53$ GeV \cite{gj} where $\rho$ and $\omega$ contributions to the $\pi$ spectra exceed 60\% \cite{kd}. 
At LHC not only do vector and tensor mesons  decay into pions, but various baryon resonances can also contribute to additional pions.  Since resonance contribution is also present at RHIC, but not considered in Ref.\ \cite{hy2}, we put aside the region at $p_T<1$ GeV/c so as to give emphasis to the difference between what we can calculate at LHC and RHIC for $p_T>1$ GeV/c.

What is notable in these spectra is the important role that the shower parton $S$ plays throughout the $\pt$ range shown. In all of them $TS$ and $TTS$ components are of the same magnitudes as $TT$ and $TTT$, or higher. In fact, for $\pt>0.5$ GeV/c,  $TS$ is larger than $TT$, and $TTS$ is larger than $TTT$ for all hadrons produced. It means that at LHC minijets are pervasive and their effects dominate the spectra throughout the low and intermediate $\pt$ regions. For $\pt<5$ GeV/c, $SS, TSS$ and $SSS$ components are all negligible. Thus the traditional fragmentation of jets can be ignored. In the recombination approach to hadronization $TS$ and $TTS$ components represent the medium effect on semihard partons, which lose energy to the medium and then regain some of the momenta back by coalescing with enhanced thermal partons to form hadrons. That they are important for $\pt>3$ GeV/c is expected, as has been found in the study of RHIC physics \cite{hy2}, but to see them as being so important for $\pt>1$ GeV/c is a new revelation at LHC.

\begin{figure}[tbph]
\centering
\vspace*{-2cm}
\includegraphics[width=.9\textwidth]{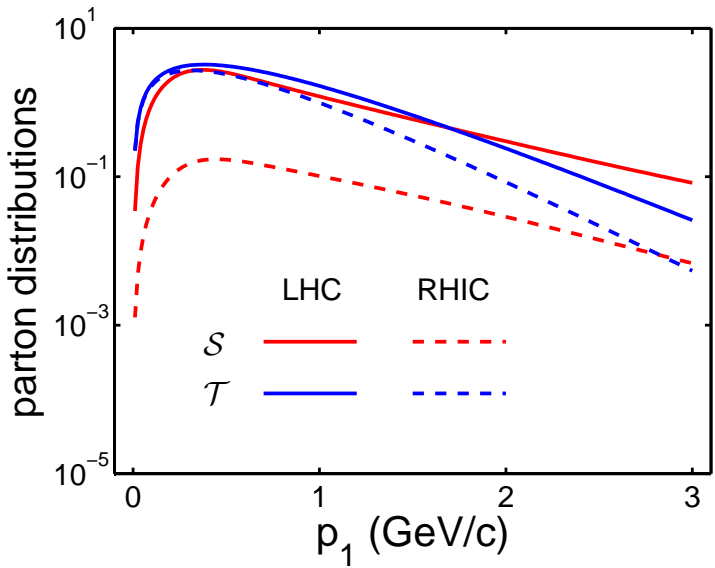}
\vspace*{-11.5cm}
\caption{(Color online) Shower and thermal parton \dis s at LHC and RHIC. }
\end{figure}
\begin{figure}[tbph]
\centering
\includegraphics[width=.65\textwidth]{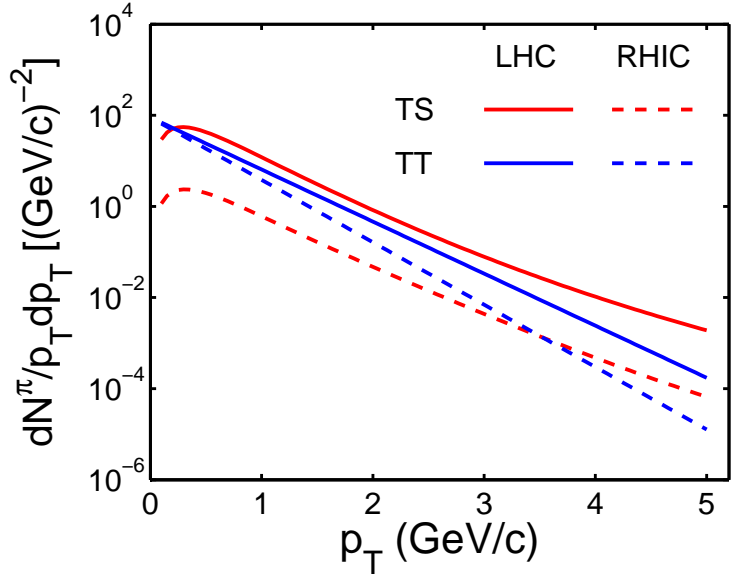}
\caption{(Color online) Transverse momentum \dis s of $\pi$ at LHC and RHIC.}
\end{figure}

To help to understand more clearly how hadron production at LHC differs from that at RHIC, we show in Fig.\ 5 the shower and thermal parton \dis s at LHC and RHIC. ${\cal S}$ and ${\cal T}$ are as defined in Eqs.\ (B6) for gluon jet and in (5), and shown by red and blue lines, respectively. Appendix B contains the details about $\cal S$ at LHC, which is more than an order of magnitude higher than ${\cal S}$ for RHIC. On the other hand, ${\cal T}$ is the same at low $p_1$ for both energies, but the exponential decrease at intermediate $p_1$ is steeper at RHIC ($T=0.32$ GeV) than at LHC ($T=0.38$ GeV). As a consequence, ${\cal T}$ at RHIC is much larger than ${\cal S}$ at $p_1<2$ GeV/c and they cross over at around $p_1=3$ GeV/c, whereas ${\cal S}$ and ${\cal T}$ at LHC are approximately the same for $p_1<2$ GeV/c. These differences have significant effects on the pion spectra. In Fig.\ 6 we compare the $p_T$ \dis s of pions arising from TS and TT recombination at the two energies. Evidently, TT dominates over TS at $p_T<3$ GeV/c at RHIC, but TS  at LHC is larger at all $p_T$. It is then clear how minijets are so important at LHC that the pion spectrum become much harder than at RHIC, and it is not because of enhanced flow.

As shown in Appendix A, there are many terms in our calculation. To get a good fit of all four spectra with the adjustment of essentially only two parameters, $T$ and $\kappa$, is a significant achievement that provides a transparent picture of the relative importance of the various thermal and shower partons at $\pt<5$ GeV/c. The quality of our fits can be seen from Figs.\ 7 (a) and (b) where the particle ratios $p/\pi$ and $\Lambda/K$ are shown. 
For the former the excellent agreement with data is partially due to the capability of our formalism to reproduce the $p$ and $\pi$ spectra accurately and separately for $\pt>1.5$ GeV/c, and partially because we have included the resonance contribution to the pions at $\pt<1.5$ GeV/c that we did not calculate. Since the reliable portion of our calculation is in the intermediate $\pt$ region above 1.5 GeV/c, the peaking of the $p/\pi$ ratio at $\pt\approx 3$ GeV/c and the gentle falloff above that are our prediction.
The $\Lambda/K$ ratio in Fig.\ 7(b) does not show perfect agreement, but it is in linear scale that amplifies the deviations of the calculated spectra from the data, which are seen in Figs.\ 2 and 4 to be well reproduced in log scale. From that more tolerant point of view the general agreement of the calculated ratio with data may be regarded as remarkably good.
The qualitative notion gained from the data that baryons are produced with more transverse momentum \cite{fa} is now given a quantitative interpretation that shower partons from minijets harden the $\pt$ spectra because of the larger phase space opened up at higher $\pt$ when three quarks recombine, one of which being the harder shower parton (as can be seen in Fig.\ 10 in Appendix B).

\begin{figure}[tbph]
\centering
\hspace*{-8.5cm}
\includegraphics[width=0.5\textwidth]{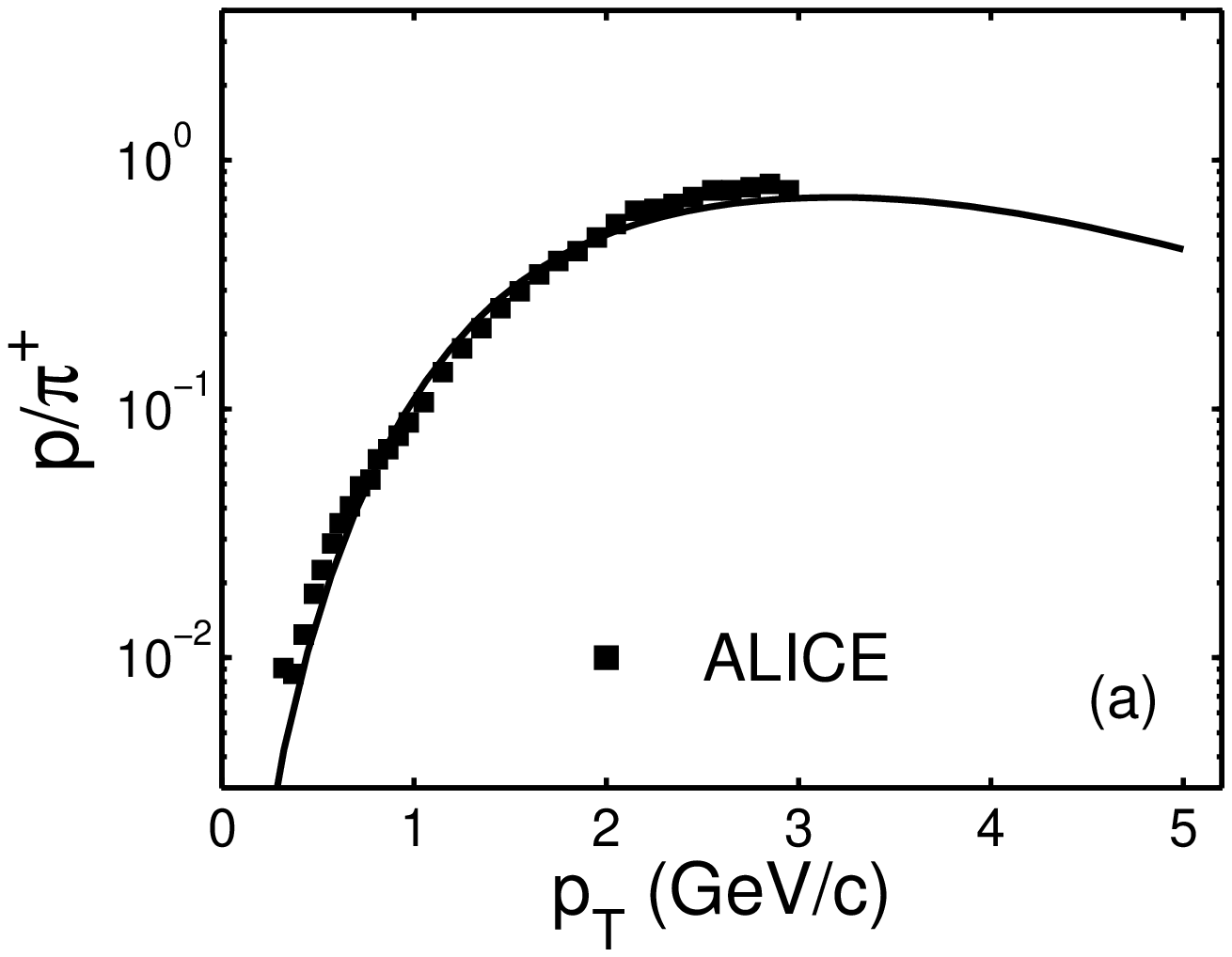}

\centering
\vspace*{-6.2cm}
\hspace*{8.5cm}
\includegraphics[width=0.5\textwidth]{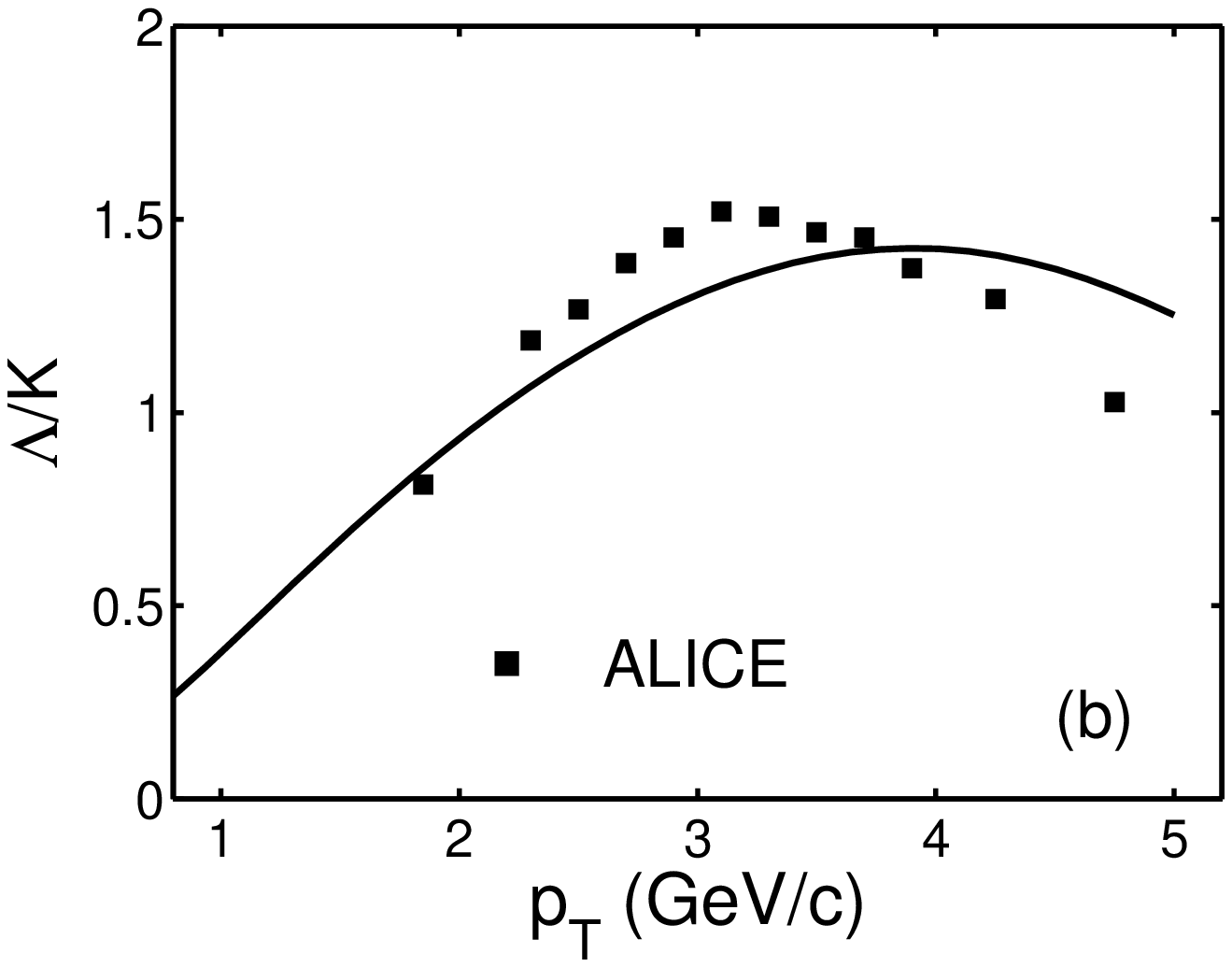}
\caption{Particle ratios of (a) $p/\pi$ and (b) $\Lambda/K$. The data are from Ref.\ \cite{mf,ib} for centrality $0-5\%$.}
\end{figure}

ALICE also has data on the $\pt$ \dis\ of all charged particles \cite{ka}. If we regard $\Lambda^0$ as representative of $\Sigma^+$, then adding  our result of $\pi, K, p$ and $\Lambda$ should come close to all charges. In Fig.\ 8 we show the sum of those four by the black line, which almost saturates the data, leaving very little room for baryons that are not included. The plot ends at $\pt\sim 5$ GeV/c where our calculation ends, but the data go on to $\pt\sim 15$ GeV/c. The mismatch will be large for $\pt>5$ GeV/c due to aspects of jet physics that we have not taken into account in the present study, but will be examined in the future.

\begin{figure}[tbph]
\centering
\includegraphics[width=0.6\textwidth]{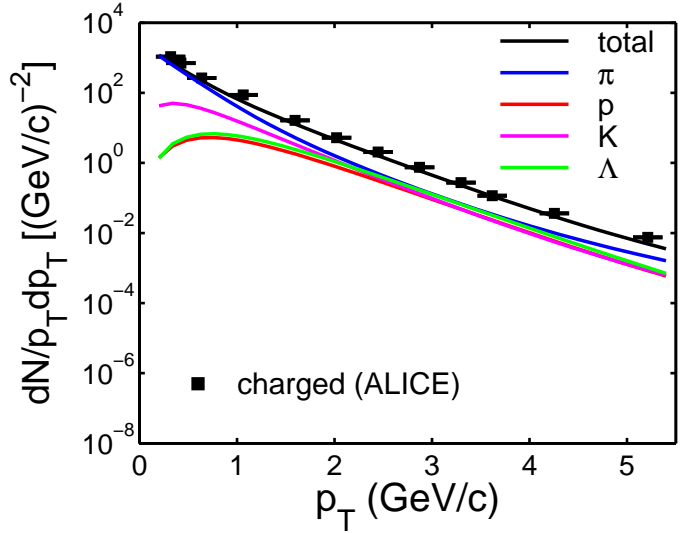}
\caption{(Color online) Charged particle $\pt$ \dis\ in black line that is the sum of the four components $\pi$ (red), $p$ (purple), $K$ (blue), and $\Lambda$ (green). The data are from Ref.\ \cite{ka} for centrality 0-5\%.}
\end{figure}

The value of $T$ at 0.38 GeV is slightly larger than 0.32 GeV that was determined for thermal partons at RHIC \cite{hy2}. That is eminently reasonable for the hotter plasma created at LHC. Since the thermal partons that participate in hadronization are at late time, $T$ is not the temperature of the system at initial equilibration time, which would be much higher. After expansion and cooling the value at LHC is, nevertheless, still higher than at RHIC because there are many more minijets produced, which lose energy to the medium and raise the ambient $T$.

The value of $\kappa$ at 2.6 implies that on average roughly $1-\kappa^{-1}=60\%$ of the initial parton energy is lost to the medium. Thus  the momenta of the  hard and semihard partons created at LHC suffer severe degradation as they traverse the dense medium. The precise value of $\kappa$ may change at higher $\pt$ since the degree of momentum degradation can depend on the initial parton momentum.

\section{CONCLUSION}

We have studied the properties of hadron production in the difficult intermediate $\pt$ region that is too low for reliable calculation in pQCD, and too high for hydrodynamical description. Although we have been successful in reproducing the data, our main point is, however, not so much on the accuracy of the fits of the data, but on learning about the physics at LHC. Since our treatment of the problem is focused only on hadronization at the final stage of the evolution of the partonic system, we cannot address issues related to the dynamical development in time. Nevertheless, the investigation provides quantitative  information about minijets that are crucial in the  understanding of the nature of hadronic spectra. A distinctive feature about our approach is that the production of mesons and baryons with widely different masses can be described on the basis of a common system of partons without any specific reference to flow. In the intermediate $\pt$ region between 2 and 5 GeV/c, the non-flow part of the hydrodynamical description can presumably be related to the thermal-shower component of recombination. Hadron masses do not enter explicitly in our calculation, although they are implicitly involved in the determination of the recombination functions. The relative magnitudes of the TT, TS and SS components for mesons (and of TTT, TTS, TSS, and SSS components for baryons) provide a clear picture of the smooth transition throughout that $\pt$ region, and unify the very different hadronic sectors.
 That picture offers an illuminating complement to the conventional hydrodynamical description without contradicting it.
 
 While the shower partons from minijets play the important role in hadronization in the intermediate $\pt$ region, harder jets will, of course, become more important at higher $\pt$. We do not expect our description to change significantly when we extend our consideration to that region, except for the need to study also multi-jet recombination. Before doing that, we still have to examine the centrality dependence in the intermediate region. That subject is interesting not only because of the data on the second harmonic, $v_2(\pt)$, that is usually interpreted as elliptic flow, but especially because  the dominance of minijets implies the importance of non-flow. The $\phi$ dependence of momentum degradation at RHIC has been considered in our approach previously \cite{hy3}, and can be adapted for collisions at LHC. That will provide another test of the validity of our finding here about the minijets.

\section*{Acknowledgment}

This work was supported,  in part,  by the U.\ S.\ Department of Energy under Grant No. DE-FG02-96ER40972 and by the Scientific Research Foundation
for Young Teachers, Sichuan University under No. 2010SCU11090 and
Key Laboratory of Quark and Lepton Physics under Grant No. QLPL2009P01.

 \newpage
 
 \begin{appendix}
 \section{Hadronic Distributions for $\pi, K, p$ and $\Lambda$}
 
 We summarize in this Appendix the formulas for all the hadronic \dis s. They follow from the general equations described in Sec. II and the specific recombination functions for the four hadrons. We shall leave out the multi-jet (2j and 3j) contributions because they are small and negligible for $\pt<5$ GeV/c, but we exhibit explicitly the various $\cal T$ and $\cal S$ combinations. As explained in Sec. IV we approximate the sum over all hard parton species
 by the dominant gluon jet contribution multiplied by a constant factor $\sigma$ (set to be $\sigma=1.2$) that roughly accounts for the combined contribution of all other quark-jet terms. That significantly simplifies the calculation without compromising the general features of the results.
 
 \subsection{Pion}
 \begin{eqnarray}
{dN_\pi^{TT}\over \pt d\pt}&=&{C^2\over 6} e^{-\pt/T} ,  \\
{dN_\pi^{TS}\over \pt d\pt}&=&{2C\over \pt^3} \int_0^{\pt} dp_1 p_1 e^{-p_1/T} {\cal S}^q(\pt-p_1,\bar\xi) ,  \\
{dN_\pi^{SS}\over \pt d\pt}&=&{1\over \pt} \int {dq\over q^2} \sigma\bar F_g(q,\bar\xi) {D'}_g^{\pi}(\pt,q) ,
\end{eqnarray}
 where
\bq
{\cal S}^q(p_2,\xb)=\int{dq\over q}\sigma\bar F_g(q,\xb) {S'}_g^q(p_2,q)
\eq
The variable $q$ is the momentum of gluon jet at the medium surface, while the index $q$ refers to light quarks to be distinguished from $s$ quark whose role in the RF of kaon is different. The SPD ${S'}_g^q$ and FF ${D'}_g^{\pi}$ are modified at low momentum in ways discussed  in Appendix B.

\subsection{Kaon}
\bq
{dN_K^{TT}\over \pt d\pt}&=&N_K B(a+2,b+2){C^2\pt\over m_T^K} e^{-\pt/T} ,  \\
{dN_K^{TS}\over \pt d\pt}&=&{N_KC\over m_T^K \pt^{a+b+2}} \int_0^{\pt} dp_1 p_1^a(\pt-p_1)^b \nonumber \\
&&\times  [p_1e^{-p_1/T} {\cal S}^s(\pt-p_1,\bar\xi) + (\pt-p_1) e^{-(\pt-p_1)/T} {\cal S}^q(p_1,\xb)],  \\
{dN_K^{SS}\over \pt d\pt}&=&{1\over m_T^K} \int {dq\over q^2} \sigma\bar F_g(q,\bar\xi) {D'}_g^{K}(\pt,q) ,
\end{eqnarray}
 where ${\cal S}^s(p_2,\xb)$ is as defined in Eq.\ (A4), except for ${S'}_g^s(p_2,q)$ replacing 
 ${S'}_g^q(p_2,q)$. The RF for kaon is found in Ref.\ \cite{hy6} to be for $a=1$ and $b=2$.
 
 \subsection{Proton}
\bq
{dN_p^{TTT}\over \pt d\pt}&=&g_{st}^pN_p B(\alpha+2,\beta+2)B(\alpha+2,\alpha+\beta+4){C^3\pt^2\over m_T^p} e^{-\pt/T} ,  \\
{dN_p^{TTS}\over \pt d\pt}&=&{g_{st}^pN_p 2C^2\over m_T^p \pt^{2\alpha+\beta+3}} \int_0^{\pt} dp_1 \int_0^{\pt-p_1} dp_2 (p_1p_2)^{\alpha+1}(\pt-p_1-p_2)^{\beta} \nonumber \\
	&& \hspace{2cm} \times e^{-(p_1+p_2)/T} {\cal S}^q(\pt-p_1-p_2,\bar\xi),  \\
{dN_p^{TSS}\over \pt d\pt}&=&{g_{st}^pN_p 2C\over m_T^p \pt^{2\alpha+\beta+3}} \int_0^{\pt} dp_1 \int_0^{\pt-p_1} dp_2 (p_1p_2)^{\alpha}(\pt-p_1-p_2)^{\beta}  \nonumber \\
	&& \hspace{2cm} \times p_1e^{-p_1/T} {\cal S}^{qq}(p_2,\pt-p_1-p_2,\bar\xi), \\
{dN_p^{SSS}\over \pt d\pt}&=&{1\over m_T^p} \int {dq\over q^2} \sigma\bar F_g(q,\bar\xi) {D'}_g^{p}(\pt,q) ,
\end{eqnarray}
 where
 \bq
 {\cal S}^{qq}(p_2,p_3,\xb)=\int{dq\over q}\sigma\bar F_g(q,\xb) {S'}_g^q(p_2,q){S'}_g^q(p_3,q-p_2).
 \eq
 The factor $g_{st}^p$ is 1/12 when feeddown from $\Delta$ is excluded \cite{hy}. The RF parameters for proton are $\alpha=1.75$ and $\beta=1.05$ \cite{hy5}.
 
  \subsection{$\Lambda$}
\bq
{dN_\Lambda^{TTT}\over \pt d\pt}&=&g_{st}^\Lambda N_\Lambda B(\alpha+2,\beta+2)B(\alpha+2,\alpha+\beta+4){C^3\pt^2\over m_T^\Lambda} e^{-\pt/T} ,  \\
{dN_\Lambda^{TTS}\over \pt d\pt}&=&{g_{st}^\Lambda N_\Lambda C^2\over m_T^\Lambda \pt^{2\alpha+\beta+3}} \int_0^{\pt} dp_1 \int_0^{\pt-p_1} dp_2 (p_1p_2)^{\alpha+1}(\pt-p_1-p_2)^{\beta} \nonumber \\
	&& \hspace{1cm} \times e^{-(p_1+p_2)/T} [2{\cal S}^q(\pt-p_1-p_2,\bar\xi)+{\cal S}^s(\pt-p_1-p_2,\bar\xi)],  \\
{dN_\Lambda^{TSS}\over \pt d\pt}&=&{g_{st}^\Lambda N_\Lambda C\over m_T^\Lambda \pt^{2\alpha+\beta+3}} \int_0^{\pt} dp_1 \int_0^{\pt-p_1} dp_2 (p_1p_2)^{\alpha}(\pt-p_1-p_2)^{\beta}  \nonumber \\
	&& \hspace{1cm} \times p_1e^{-p_1/T}[2{\cal S}^{qq}(p_2,\pt-p_1-p_2,\bar\xi)+{\cal S}^{qs}(p_2,\pt-p_1-p_2,\bar\xi)], \\
{dN_\Lambda^{SSS}\over \pt d\pt}&=&{1\over m_T^\Lambda} \int {dq\over q^2} \sigma\bar F_g(q,\bar\xi) {D'}_g^{\Lambda}(\pt,q) ,
\end{eqnarray}
where $g_{st}^\Lambda=1/8$ (1/2 for $\Lambda^0$ or $\Sigma^0$, and 2/8 from spin consideration). The RF parameters for $\Lambda$ are $\alpha=1, \beta=2$ \cite{hy6}. ${\cal S}^{qs}(p_2,p_3,\bar\xi)$ is as defined in Eq.\ (A12), but with ${S'}_g^s(p_3)$ replacing ${S'}_g^q(p_3)$.

  \section{Shower Parton Distribution}

We derive in this Appendix the shower parton \dis\ we use for the study of hadron production at LHC. The basic idea is already described in Refs.\
\cite{hy1,hy4}; only the parametrization is now different. Fragmentation function (FF) at high $Q^2$ in deep inelastic scattering gives the hadron
\dis\ in a quark or gluon jet. It does not specify the way in which hadrons are formed. In the recombination model FF is described as a two-step
process, first the development of shower partons in a jet, then the coalescence of the shower partons to form a hadron. For pion production it is as
expressed in Eq.\ (\ref{7}), which should be augmented by the dependence of $D_i^{\pi}(x,\mu^2)$ and $S_i^j(x,\mu^2)$ on the energy scale $\mu$. Since
the evolution of $D_i^{\pi}(x,\mu^2)$ in $\mu^2$ can be tracked experimentally and theoretically \cite{kkp}, the $\mu^2$ dependence of
$S_i^j(x,\mu^2)$ can be determined accordingly by use of Eq.\ (\ref{7}). However, to include that dependence in the application of $S_i^j(x)$ in Eq.\
(\ref{12}) for heavy-ion collisions is too complicated and more meticulous than necessary in view
of the many other unavoidable approximations. Thus a fixed $\mu^2$ is used in practice. For RHIC we have used $\mu=10$ GeV \cite{hy1,hy4}. Now for LHC
we continue to use the same $\mu$, since the hadronization scale at late time is the same. However, we improve the determination of the SPD and
include scale-breaking effects due to a cut-off at low $p_1$.
\begin{figure}[tbph]
\includegraphics[width=0.6\textwidth]{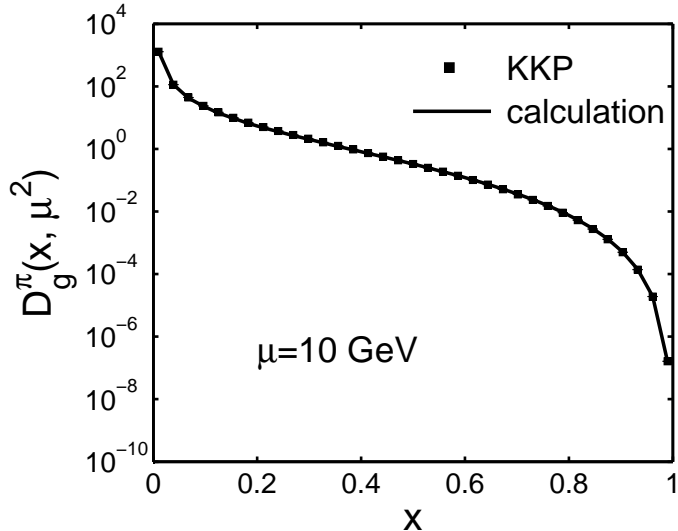}\\
\begin{flushleft}
\caption{Fit of KKP fragmentation function $D_g^{\pi}(x,\mu^2)$ \cite{kkp} (squares) by Eq.\ (B1) using the recombination formulas Eqs.\ (7) and (27), as shown by the solid line.}
\end{flushleft}
\end{figure}

The gluon FF $D_g^{\pi}(x,\mu^2)$ for $\mu=10$ GeV is shown by the square points in Fig.\ B-1,  following the parametrization given in Ref.\ \cite{kkp}.
To reproduce that $x$ dependence, we use the SPD from gluon to light quark $q$
\begin{equation}
{S}_g^q(x)=ax^b(1-x)^c(1+dx^e) \label{B1}
\end{equation}
where $a=0.739, b=-0.28, c=4.387, d=4.502$ and $e=10.469$.
The solid line in Fig.\ 9 shows the result of our calculation based on Eqs.\ (\ref{7}) and (\ref{B1}). The fit is evidently very good.

In the application of ${S}_g^q(x)$ to hadronization processes in heavy-ion collisions, we have discussed in Sec.\ IV how the shower partons at very low
momenta $p_i$ are merged into the region dominated by thermal partons. 
The semihard partons lose energy to the medium whose thermal partons are enhanced in such a way as  to make the distinction between thermal and shower partons meaningless.
The peaking of the SPD at very low $p_i$ is therefore unrealistic. Since the inverse slope of the thermal parton \dis\ is to be determined phenomenologically that includes the effect of energy loss of the semihard partons, we require that the SPD should not exceed the exponential behavior of the thermal partons.
The scale invariant thermal \dis\ ${\cal T}(p_1)$ given in Eq.\ (\ref{5}) is shown in Fig.\ 10 by the red line; it dips at small $p_1$
because of the prefactor $Cp_1$. 
The integrated shower \dis\ ${\cal S}_g^q(p_i)$, defined as in Eq.\ (\ref{15}) but for gluon jet ($i=g$) only and with $\bar\xi$ replaced by $\kappa$ in $\bar F_g(q,\kappa)$ given in Eq.\ (\ref{38}), is shown as a function of $p_1$   by the dashed line Fig.\ 10. We see that
${\cal S}_g^q(p_1)$ exhibits a  power-law behavior   for $p_1>1$ GeV/c that is expected, and  is larger than the exponential behavior of ${\cal T}(p_1)$. However, the peaking at low $p_1$
for $p_1<0.5$ GeV/c is due to the unreliability of the method of determining $S_g^q(x)$ at low $x$. Since for physical reasons we want the thermal
partons to dominate at low $p_1$, we introduce a cut-off factor on the SPD.
A cut-off with a particular scale implies breaking of scale-invariance described by $x$. Such a breaking is reasonable in low-$q$ processes. If we
write $x=p_T/q$ in $D_g^{\pi}(x)$, but now in the non-scaling form $D_g^{\pi}(p_T,q)$, we introduce the low-$p_T$ cut-off as follows
\begin{equation}
{D'}^{\pi}_g(p_T,q)=D^{\pi}_g(p_T,q)\gamma_1(p_T),   \qquad \gamma_1(p_T)=1-e^{-p_T^2}.  \label {B2}
\end{equation}
The shower parton \dis\ must therefore also be modified in accordance to
\begin{equation}
{p_T\over q}{D'}^{\pi}_g(p_T,q)={1\over p_T} \int_0^{p_T} dp_1 {S'}^{q}_g(p_1,q){S'}^{q}_g(p_T-p_1,q-p_1) ,   \label {B3}
\end{equation}
where Eq.\ (\ref{27}) is used in (\ref{7}) with $x_1=p_1/q$. The modified ${S'}^{q}_g(p_1,q)$ is now
\begin{equation}
{S'}^{q}_g(p_1,q)=S^q_g(p_1/q) \gamma_2(p_1) , \label {B4}
\end{equation}
where the corresponding cut-off in $p_1$ is
\begin{equation}
\gamma_2(p_1)=1-e^{-(p_1/0.3)^2} . \label {B5}
\end{equation}
\begin{figure}[tbph]
\includegraphics[width=0.8\textwidth]{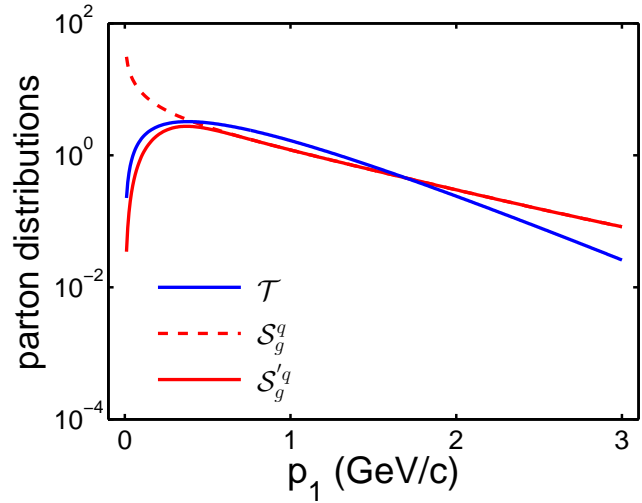}\\
\vspace*{-10.cm}
\begin{flushleft}
\caption{(Color online) Thermal \dis\ ${\cal T}(p_1)$ is depicted by the blue line for $T=0.38$ GeV. Shower parton distributions are shown by ${\cal
S}^q_g(p_1)$ (dashed line) and ${{\cal S}'}^q_g(p_1)$ (solid red line); the latter includes the low-$p_1$ cut-off.}
\end{flushleft}
\end{figure}
It can be demonstrated that the two damping factors $\gamma_1(p_T)$ and $\gamma_2(p_1)$ are coordinated to satisfy Eq.\ (\ref{B3}) for all $p_T$ in the
range of $1<p_T<15$ GeV/c.

The consequence of this cut-off on the  integrated shower \dis\
\bq
{\cal S'}_g^q(p_1,\kappa)=\int {dq\over q}\bar F_g(q,\kappa) {S'}_g^q(p_1,q) 
\eq
 is shown by the solid (black) line in Fig.\ 10. Note that it is now just lower than the red line of the thermal \dis\ ${\cal T}(p_1)$ for $p_1<0.5$ GeV/c,which is the criterion of this cut-off. In actual computation of the hadronic spectra this modified \dis\ $\cal S'$ is used in all shower \dis s generically expressed as $\cal S$ in Sec.\ II and Appendix A.

 \end{appendix}

\end{document}